\documentclass[12pt,a4paper,notitlepage,fleqn]{article}

\usepackage{setspace}
\usepackage[margin=1in]{geometry}
\usepackage[symbol]{footmisc}

\usepackage[T1]{fontenc}
\usepackage[utf8]{inputenc}
\usepackage{authblk}

\makeatletter

\newcommand{\fboxsubsec}[1]{
	\begin{flushleft}
		#1
	\end{flushleft}
	}
\renewcommand{\subsection}{\@startsection{subsection}{2}{0pt}
	{1ex}
	{0.5ex}
	{\reset@font\it\fboxsubsec}
	}
\makeatother

\title{Examining the Dynamic Asset Market Linkages\\
under the COVID-19 Global Pandemic}% 

\author{Akihiko Noda$^{a,b}$\thanks{\scriptsize Corresponding Author. E-mail: anoda@meiji.ac.jp, Tel. +81-3-3296-2265, Fax. +81-3-3296-4347}

{\scriptsize ${}^{a}$ \it School of Commerce, Meiji University, 1-1 Kanda-Surugadai, Chiyoda-ku, Tokyo 101-8301, Japan}

{\scriptsize ${}^{b}$ \it Keio Economic Observatory, Keio University, 2-15-45 Mita, Minato-ku, Tokyo 108-8345, Japan}}

\date{This Version: \today}

\renewcommand\thefootnote{\arabic{footnote}}

\usepackage{graphicx}
\usepackage[]{natbib}%
\usepackage{amsmath,amssymb}%
\usepackage{ascmac}%
\usepackage{multirow}%
\usepackage{lscape}%
\usepackage{subfigmat}

\usepackage{pifont}%
\usepackage{arydshln}%
\usepackage[format=hang]{caption}
\usepackage[all]{xy}
\usepackage{url}
\bibpunct{(}{)}{;}{a}{}{,}
\def\hsymbu#1{\smash{\lower1.7ex\hbox{\huge$#1$}}}

\def\ve #1{{\mbox{\boldmath $#1$}}}

\newcommand{\citetapos}[1]{\citeauthor{#1}'s \citeyearpar{#1}}
\newcommand{\citeapos}[2]{\citeauthor{#1}'s (\citeyear{#2})}

\def\ve #1{{\mbox{\boldmath $#1$}}}

\begin{document}

%\onehalfspacing

\begin{titlepage}

\renewcommand{\thepage}{}
\renewcommand{\thefootnote}{\fnsymbol{footnote}}

\maketitle

\vspace{-10mm}

%\begin{quote}
\noindent
\hrulefill

{\bfseries Abstract:} This study examines the dynamic asset market linkages under the COVID-19 global pandemic based on market efficiency, in the sense of \citet{fama1970ecm}. Particularly, we estimate the joint degree of market efficiency by applying \citeapos{ito2014ism}{ito2014ism,ito2017aae} Generalized Least Squares-based time-varying vector autoregression model. The empirical results show that (1) the joint degree of market efficiency changes widely over time, as shown in \citetapos{lo2004amh} adaptive market hypothesis, (2) the COVID-19 pandemic may eliminate arbitrage and improve market efficiency through enhanced linkages between the asset markets; and (3) the market efficiency has continued to decline due to the Bitcoin bubble that emerged at the end of 2020.\\

\noindent
{\bfseries Keywords:} COVID-19; Asset Market Linkages; Adaptive Market Hypothesis; Efficient Market Hypothesis; GLS-Based Time-Varying Model Approach.\\

\noindent
{\bfseries JEL Classification Numbers:} G12; G14.

\noindent
\hrulefill
%\end{quote}

\end{titlepage}

\bibliographystyle{asa}%\

%\pagebreak

\section{Introduction}\label{sec:asset_linkages_intro}
Since \citet{nakamoto2008btc} establishment Bitcoin as a cryptocurrency system, its market capitalization has expanded and experienced several price bubbles. A particularly serious price bubble has persisted since 2020, and the market capitalization of Bitcoin reached approximately USD 1.2 trillion on April 14, 2021\footnote{The historical data for total market capitalization are available on the webpage of CoinMarketCap (\url{https://coinmarketcap.com/charts/}).}. This means that Bitcoin's market capitalization has expanded more than 10 times that before March 11, 2020, when the World Health Organization (WHO) declared COVID-19 a global pandemic. The changes in market capitalization suggest that investors regard Bitcoin as an asset, but this does not necessarily mean that they do not regard it as, for example, a safe haven. Therefore, economists have been examining the Bitcoin market from two perspectives: (i) whether it is efficient in the sense of \citet{fama1970ecm} and (ii) whether Bitcoin has the ability to hedge against other assets in the sense of \citet{baur2010igh}.

The first perspective follows \citet{urquhart2016tib}, who examines whether the price of Bitcoin is efficiently determined. In this context, several studies have examined whether \citetapos{lo2004amh} adaptive market hypothesis (AMH) holds in the Bitcoin market and have shown that market efficiency varies over time\footnote{The AMH was proposed as an alternative to \citetapos{fama1970ecm} efficient market hypothesis, which explains market efficiency (or return predictability) changes over time.}. For example, using a conventional statistical test with a rolling-window framework proposed by \citet{dominguez2003tmd}, \citet{khuntia2018amh} and \citet{chu2019amh} show that the AMH holds in the Bitcoin market. \citet{noda2021oec} directly measures the degree of market efficiency using \citeapos{ito2016eme}{ito2016eme,ito2017aae} Generalized Least Squares (GLS)-based time-varying--autoregressive (TV--AR) model and finds that the market efficiency in the Bitcoin market changes with time.

The second perspective follows \citet{klein2018bng}, who argue that Bitcoin has no stable hedging capabilities and is not a new alternative to gold in terms of being a safe heaven. However, whether bitcoin is a safe haven has been a controversy in more recent studies. \citet{conlon2020shr} show that Bitcoin does not work as a safe-haven asset against the U.S. stock market index, the S\&P 500. This implies that there are dynamic correlations between Bitcoin and traditional markets, as shown in \citet{corbet2020cec}. Contrarily, using monthly data, \citet{chan2019hbl} reveal that Bitcoin has a strong hedge for several stock market indices, and \citet{pal2019hbo} find that gold has strong hedging abilities against Bitcoin.

Thus, the Bitcoin market's efficiency and its hedging abilities against other assets have been examined separately in previous studies. However, consider \citetapos{fama1970ecm} definition of efficient market: If prices are informationally efficient, then there are no arbitrage opportunities among financial assets, and if there are no arbitrage opportunities, then asset market linkages should be confirmed. The existence of linkages implies that there are no hedging abilities (as a safe haven) to reduce financial risk in periods of financial instability. Therefore, market efficiency and hedging abilities against other assets in the Bitcoin market are research issues that should be examined simultaneously. Additionally, we should consider the possibility that market efficiency and hedging abilities in the Bitcoin market vary with time, as discussed in \citet{akhtaruzzaman2021ghs} and \citet{noda2021oec}.

This study examines the dynamics of the market linkage between the Bitcoin market and other asset markets based on market efficiency in the sense of \citet{fama1970ecm}. Specifically, we focus on the market linkages of three financial asset markets (S\&P500, Bitcoin, and gold), which have been examined in previous studies, and measure the joint degree of market efficiency based on \citeapos{ito2014ism}{ito2014ism,ito2017aae} GLS-based time-varying vector autoregressive (TV-VAR) model. The estimated time-varying joint degree of market efficiency may not only reveal changes in market linkages among the three markets, but also elucidate the time-varying hedging abilities in the three markets. Finally, we discuss what events in financial asset markets cause fluctuations in market efficiency and market linkages.

The remainder of this paper is organized as follows. Section \ref{sec:asset_linkages_model} presents our method for studying market efficiency variation over time based on \citeapos{ito2014ism}{ito2014ism,ito2017aae} GLS-based TV-VAR model. Section \ref{sec:asset_linkages_data} describes the daily prices in the three asset markets (S\&P500, Bitcoin, and gold) to calculate the returns and presents preliminary unit root test results. Section \ref{sec:asset_linkages_emp} presents our empirical results and discusses the time-varying nature of market linkages from the viewpoint of the AMH, which indicates that market efficiency changes over time. Section \ref{sec:asset_linkages_cr} concludes the paper.

%\pagebreak

\section{Model}\label{sec:asset_linkages_model}
In this section, we introduce \citeapos{ito2014ism}{ito2014ism,ito2017aae} GLS-based TV-VAR model to investigate the dynamics of asset market linkages between three securities (S\&P500, Bitcoin, and gold) from an informational efficiency perspective. Suppose that $\ve{p}_t$ is a price vector of the three securities in period $t$. Our main focus is reduced to the following conditions. 
\begin{equation}
 \mathbb E\left[ \ve{x}_t \mid \mathcal{I}_{t-1}\right]=0,\label{EMH}
\end{equation}%
where $\ve{x}_{t}$ denotes a return vector of the securities in period $t$; that is, the $i$-th component of $\ve{x}_{t}$ is $\ln {p_{i,t}}-\ln p_{i,t-1}$ for $i=S,B,G$. In other words, all expected returns in period $t$ are zero, given the information set available in period $t-1$.

When $\ve{x}_{t}$ is stationary, the Wold decomposition allows us to regard the time-series process of $\ve{x}_{t}$ as 
\begin{equation*}
 \ve{x}_{t}=\ve\mu+\Phi_{0}\ve{u}_{t}+\Phi_{1}\ve{u}_{t-1}+\Phi_{2}\ve{u}_{t-2}+\cdots,
\end{equation*}%
where $\ve\mu$ is a vector of the mean of $\ve{x}_t$ and $\left\{\ve{u}_{t}\right\} $ follows an independent and identically distributed multivariate process with a mean of zero vector, and the covariance matrix of $\sigma^{2} I$, $\sum_{i=0}^{\infty}||\Phi_{i}^\prime\Phi_{i}||<\infty$, where $\Phi_{0}=I$. Note that the efficient market hypothesis (EMH) holds if and only if $\Phi_i=0$ for all $i>0$, which suggests that how the market deviates from the efficient market reflects the impulse response, a series of $\{\ve{u}_{t}\}$. We construct an index based on the impulse response to investigate the dynamics of asset market linkages in the sense of whether the EMH holds.

The easiest way to obtain the impulse response is to use a VAR model and algebraically compute its coefficient estimates. Under some conditions, the vector return process $\{\ve{x}_t\}$ of the securities is invertible\footnote{See \citet{ito2017aae} for details.}. We consider the following TV-VAR($q$) model:
\begin{equation}
 \ve{x}_{t}=\ve\nu + A_{1}\ve{x}_{t-1}+A_{2}\ve{x}_{t-2}+\cdots+A_{q}\ve{x}_{t-q}+\ve\varepsilon_{t}; \ \ t=1,2,\ldots,T,\label{arq}
\end{equation}
where $\ve\nu$ is a vector of intercepts, $\ve\varepsilon_{t}$ is a vector of error terms with $\mathbb E\left[\ve\varepsilon_{t}\right]=\ve{0}$, \ $\mathbb E\left[\ve\varepsilon_{t}^{2}\right]=\sigma_{\ve\varepsilon}^{2} I$, and $\mathbb E\left[\ve\varepsilon_{t}\ve\varepsilon_{t-m}\right]=\ve{0}$ for all $m\neq 0$. We use \citet{ito2014ism,ito2017aae}'s idea when we measure market efficiency that varies over time. Directly applying Equation (\ref{arq}) to the model, we obtain the degree computed through the VAR estimates, $A_1,\cdots,A_q$, as follows: First, we compute the cumulative sum of the coefficient matrices of the impulse response:
\begin{equation}
 \Phi(1)=\left(I-A_1-A_2-\cdots-A_q\right)^{-1},\label{LongRunMult}
\end{equation}
Second, to measure the deviation from the efficient market, we define the joint degree of market efficiency:
\begin{equation}
 \zeta=\sqrt{\mbox{max} \left[(\Phi(1)-I)^\prime(\Phi(1)-I)\right]},\label{LongRunMult}
\end{equation} 
 Note that in the case of the efficient market, where $A_1=A_2=\cdots=A_q=0$, our degree $\zeta$ becomes zero; otherwise, $\zeta$ deviates from zero. Therefore, we call $\zeta$ the joint degree of market efficiency. When a large deviation of $\zeta$ from $0$ (both positive and negative) exists, we can regard some deviation from one as evidence of market inefficiency. If there are no arbitrage opportunities between the three markets, we consider the degree to be improving because zeta should be close to zero. In other words, the deterioration of the degree implies that market linkages weaken over time. Furthermore, we can construct a degree that varies over time in this manner when we obtain time-varying estimates in Equation (\ref{arq}).

Adopting \citeapos{ito2014ism}{ito2014ism,ito2017aae} method, we estimate the VAR coefficients in each period to obtain the degree defined in Equation (\ref{LongRunMult}) in each period. Following \citeapos{ito2014ism}{ito2014ism,ito2017aae} idea, we use a model in which all VAR coefficients except the one that corresponds to the vector of intercepts, $\ve\nu$, follow independent random walk processes. Thus, we assume that:  
\begin{equation}
 A_{l,t}=A _{l,t-1}+V_{l,t}, \ \ (l=1,2,\cdots,q), \label{rw_a}
\end{equation}
where the error term matrix $\left\{V_{l,t}\right\}$ ($l=1,2,\cdots,q$ and $t=1,2,\cdots,T$) satisfies $\mathbb E\left[V_{l,t}\right]=\ve O$ for all $t$, \ $\mathbb E\left[vec(V_{l,t})^\prime vec(V_{l,t})\right]=\sigma_v^{2} I$ and $\mathbb E\left[vec(V_{l,t})^\prime vec(V_{l,t-m})\right]=\ve O$ for all $l$ and $m\neq 0$. \citeapos{ito2014ism}{ito2014ism,ito2017aae} method allows us to estimate the TV-VAR model:
\begin{equation}
 \ve{x}_{t}=\ve\nu+A_{1,t}\ve{x}_{t-1}+A_{2,t}\ve{x}_{t-2}+\cdots+A_{q,t}\ve{x}_{t-q}+\ve\varepsilon_{t},\label{tv_arq}
\end{equation}
together with Equation (\ref{rw_a}). 

To conduct statistical inference on our time-varying degree of market efficiency, we apply a residual bootstrap technique to the TV-VAR model above. We build a set of bootstrap samples of the TV-VAR estimates under the hypothesis that all TV-VAR coefficients are zero. This procedure provides us with a (simulated) distribution of the TV-VAR coefficients, assuming that the securities return processes are generated under the EMH. Thus, we can compute the corresponding distribution of the impulse response and the degree of market efficiency. Finally, by using confidence bands derived from such a simulated distribution, we conduct statistical inference on our estimates and detect periods when the asset market linkages are weak from a market efficiency perspective.

\section{Data}\label{sec:asset_linkages_data}
We utilize the daily spot prices of S\&P500, Bitcoin, and gold (London bullion market) obtained from Yahoo!Finance (\url{https://finance.yahoo.com/}). The start and end dates for all datasets are the same: September 14, 2014 to August 31, 2021. We take the log first difference of the time series of prices to obtain the returns of each asset.
\begin{center}
(Table \ref{asset_linkages_table1} here)
\end{center}
Table~\ref{asset_linkages_table1} shows the descriptive analysis for the returns. We confirm that the mean and standard deviation of returns on Bitcoin are higher than those of other assets. This indicates that Bitcoin is a risky asset and is not an alternative to gold, which is a representative risk-free asset. For estimations, all variables that appear in the moment conditions should be stationary; we confirm whether all our variables satisfy this condition using \citetapos{elliott1996eta} augmented Dickey—Fuller--Generalized Least Squares (ADF--GLS) test. The ADF--GLS test rejects the null hypothesis that all returns contain a unit root at the 1\% significance level.

\section{Empirical Results}\label{sec:asset_linkages_emp}
\begin{figure}[h]
 \caption{Time-Varying Joint Degree of Market Efficiency}
 \label{asset_linkages_fig1}
 \begin{center}
 \includegraphics[scale=0.6]{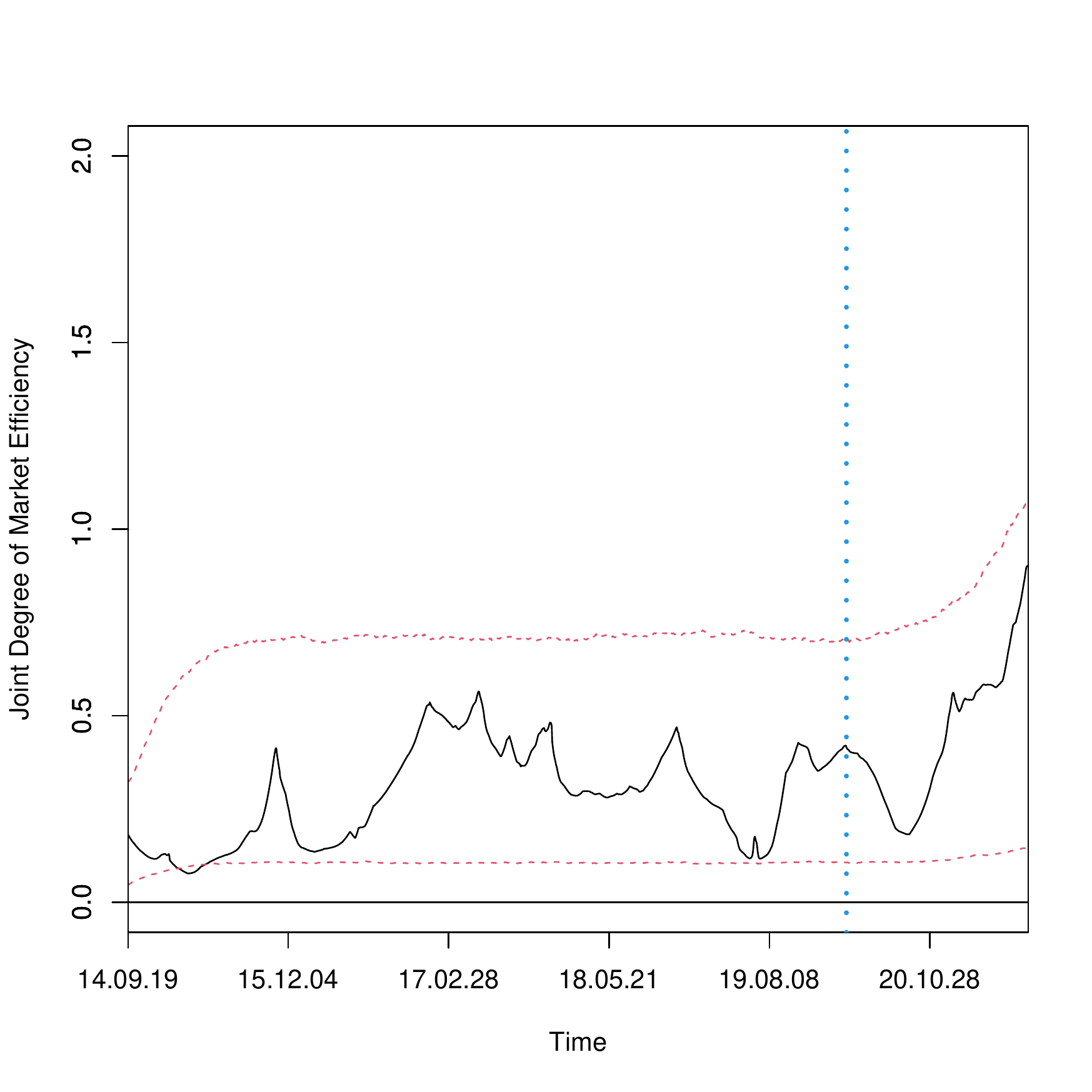}
\vspace*{5pt}
{
\begin{minipage}{420pt}
\scriptsize
\underline{Notes}:
\begin{itemize}
 \item[(1)] The dashed red lines represent the 95\% confidence intervals of the efficient market degrees.
 \item[(2)] The dotted blue line represents the time period when the World Health Organization declared the COVID-19 a ``global pandemic'' on March 11, 2020.
 \item[(2)] We ran bootstrap sampling 10,000 times to calculate the confidence intervals.
 \item[(3)] R version 4.1.1 was used to compute the estimates.
\end{itemize}
\end{minipage}}%
\end{center}
\end{figure}

We assume a standard VAR model with constants and select the optimal lag order for the model using \citetapos{schwarz1978edm} Bayesian information criterion (BIC). We therefore choose the first-order standard VAR model, hereafter called the standard VAR(1) model. Table \ref{asset_linkages_table2} presents the preliminary results for the model.
\begin{center}
(Table \ref{asset_linkages_table2} here)
\end{center}
The estimates of the autoregressive terms are not statistically significant at conventional levels, except for S\&P500. Furthermore, the null hypothesis of \citetapos{granger1969icr} causality test for the S\&P 500 is rejected at the 10\% level of statistical significance, suggesting that the S\&P 500 is causal in the sense of prediction for other variables. Thereafter, we employ \citetapos{hansen1992a} parameter constancy test to examine whether the parameters of the standard VAR(1) model are time stable. As Table \ref{asset_linkages_table2} shows, we reject the null hypothesis of constant parameters against the parameter variation as a random walk at the 1\% significance level. Therefore, we estimate the time-varying joint degree of market efficiency using \citeapos{ito2014ism}{ito2014ism,ito2017aae} GLS-based TV-VAR model to explore dynamic asset market linkages under the COVID-19 pandemic. 
\begin{center}
(Figure \ref{asset_linkages_fig1} here).
\end{center}

Figure \ref{asset_linkages_fig1} shows the variation of the joint degree of market efficiency between the three assets. We first find that the market efficiency changes widely over time as described in \citetapos{lo2004amh} AMH, and that the price formation function works well over the entire period. This fact is also consistent with \citet{ito2016eme} and \citet{noda2016amh,noda2021oec} who investigate the time-varying nature of market efficiency in the stock and cryptocurrency markets. Further, the market efficiency improves from March 11, 2020, when the WHO declared COVID-19 a global pandemic. In other words, the pandemic may have eliminated arbitrage and improved market efficiency through enhanced linkages between the asset markets. \citet{caferra2021wra} find a financial contagion in March 2020, since both cryptocurrency and stock prices fell steeply. 

Subsequently, market efficiency has continued to decline due to the Bitcoin bubble that emerged at the end of 2020. As many previous studies argue, although Bitcoin is useful for structuring a portfolio from a diversification perspective, it has poor hedging capabilities compared to other financial assets (see \citet{dyhrberg2016bgd,dyhrberg2016hcb}, \citet{bouri2017dbh,bouri2017ohs}, and \citet{baur2018bgu} for details). Additionally, cryptocurrency prices are not formed as efficiently as those of other typical financial assets such as gold, stocks, and foreign exchange, as shown in \citet{ai2018eml}. This suggests that, under the Bitcoin bubble, price formation was extremely speculative, which weakened the market linkage through a significant decrease in market efficiency. \citet{pal2019hbo} show that gold provides a better hedge against Bitcoin, but as \citet{akhtaruzzaman2021ghs} finds, the hedging function of gold as a safe asset was also impaired during the COVID-19 global pandemic.

\section{Concluding Remarks}\label{sec:asset_linkages_cr}
In this study, we examine the dynamic asset market linkages under the COVID-19 global pandemic from the market efficiency perspective in the sense of \citet{fama1970ecm}. If market efficiency changes over time, as pointed out in \citet{lo2004amh}, the existence of arbitrage opportunities should also change over time. Moreover, when arbitrage opportunities between markets do not exist (do exist), the linkages between markets are expected to be stronger (weaker). Particularly, we estimate the joint degree of market efficiency based on \citet{ito2014ism,ito2017aae}'s GLS-based TV-VAR model. The empirical results show that (1) the joint degree of market efficiency changes widely over time, as shown in Lo(2004); (2) the market efficiency improved immediately after the WHO's declaration of COVID-19 as a global pandemic on March 11, 2020, and (3) the market efficiency declined due to the Bitcoin bubble that emerged at the end of 2020, and continues to do so. Therefore, the empirical results support the AMH. Further, the pandemic may eliminate arbitrage and improve market efficiency through enhanced linkages between asset markets.

\section*{Acknowledgments}

The author would like to thank Mikio Ito, Yoichi Tsuchiya, Tatsuma Wada, and the seminar participants at Keio University and Meiji University for their helpful comments and suggestions. The author is also grateful for the financial assistance provided by the Japan Society for the Promotion of Science Grant in Aid for Scientific Research, under grant number 19K13747. All data and programs used are available upon request.

\pagebreak

\bigskip

\bigskip

\bigskip

\setcounter{table}{0}
\renewcommand{\thetable}{\arabic{table}}

%\begin{figure}[p]
% \caption{The Returns on the BTC, ETH, and XRP}
% \label{crypto_fig1}
% \begin{center}
% \includegraphics[scale=0.8]{../../fig/TS_PLOT_DES.pdf}
% \end{center}
%\end{figure}

%\pagebreak

\begin{table}[h]
\caption{Descriptive Statistics and Unit Root Tests}
\label{asset_linkages_table1}
\begin{center}
\begin{tabular}{lccccccccccccc}\hline\hline
 &  &  & Mean & SD & Min & Max &  & ADF-GLS & Lags & $\hat\phi$ &  & $\mathcal{N}$ & \\\cline{4-7}\cline{9-11}\cline{13-13}
 & $R_S$ &  & 0.0005  & 0.0115  & -0.1277  & 0.0897  &  & -10.5354  & 8 & -0.1472  &  & 1685 & \\
 & $R_B$ &  & 0.0027  & 0.0472  & -0.4647  & 0.2251  &  & -4.1933  & 10 & 0.3780  &  & 1685 & \\
 & $R_G$ &  & 0.0002  & 0.0091  & -0.0526  & 0.0513  &  & -4.4568  & 11 & 0.3102  &  & 1685 & \\\hline\hline
\end{tabular}
%\vspace*{5pt}
{
\begin{minipage}{420pt}\scriptsize
{\underline{Notes:}}
\begin{itemize}
 \item[(1)] ``$R_S$,'' ``$R_B$,'' and ``$R_G$'' denote the returns of S\&P 500, Bitcoin, and gold, respectively. 
 \item[(2)] ``ADF--GLS'' denotes the augmented Dickey--Fuller GLS test statistics, and ``Lag'' denotes the lag order selected by the Bayesian information criterion.
 \item[(3)] In computing the ADF--GLS test, a model with a time trend and constant is assumed. The critical value at the 1\% significance level for the ADF--GLS test is $-3.96$
 \item[(4)] ``$\mathcal{N}$'' denotes the number of observations.
 \item[(5)] R version 4.1.1 was used to compute the statistics.
\end{itemize}
\end{minipage}}%
\end{center}
\end{table}

\begin{table}[t]
\caption{Preliminary Estimation}
\label{asset_linkages_table2}
\begin{center}
\begin{tabular}{cccccc}\hline\hline
 &  & $R_{S,t}$ & $R_{B,t}$ & $R_{G,t}$ & \\\hline
 & \multirow{2}*{$Constant$} & $0.0006$  & $0.0028$  & $0.0003$  & \\
 &  & $[0.0003]$  & $[0.0012]$  & $[0.0002]$  & \\
 & \multirow{2}*{$R_{S,t-1}$} & $-0.1910$  & $-0.0026$  & $-0.0417$  & \\
 &  & $[0.0820]$  & $[0.1160]$  & $[0.0322]$  & \\
 & \multirow{2}*{$R_{B,t-1}$} & $-0.0057$  & $-0.0011$  & $0.0021$  & \\
 &  & $[0.0119]$  & $[0.0306]$  & $[0.0042]$  & \\
 & \multirow{2}*{$R_{G,t-1}$} & $-0.0410$  & $-0.1729$  & $0.0004$  & \\
 &  & $[0.0451]$  & $[0.1342]$  & $[0.0229]$  & \\\hline
 & $\bar{R}^2$ & $0.0389$  & $-0.0006$  & $0.0009$  & \\
 & $Granger$ & $2.2812^*$  & $0.5878$ & $1.5944$ & \\\hdashline
 & $L_C$ & \multicolumn{3}{c}{$163.5671^{***}$} & \\\hline\hline
\end{tabular}
%\vspace*{5pt}
{
\begin{minipage}{420pt}\scriptsize
{\underline{Notes:}}
\begin{itemize}
 \item[(1)] ``$R_{t-p}$,'' ``$\bar{R}^2$,'' ``$Granger$,'' and ``$L_C$'' denote the vector autoregression ($p$) estimates, the adjusted $R^2$, the $F$ statistics for \citetapos{granger1969icr} causality test, and \citetapos{hansen1992a} joint $L$ statistic with variance, respectively.
 \item[(2)] ``***'' and ``*'' indicate that the null hypothesis of each test is rejected at the 1\% and 10\% significance level, respectively.
 \item[(3)] \citetapos{newey1987sps} robust standard errors are within parentheses.
 \item[(4)] R version 4.1.1 was used to compute the statistics.
\end{itemize}
\end{minipage}}%
\end{center}
\end{table}

\end{document}